\documentclass[prb,superscriptaddress,floatfix,twocolumn,showpacs,amsfonts,amsmath,amssymb]{revtex4}
\usepackage{graphicx} 
\usepackage{dcolumn} 
\usepackage{bm} 

\begin{document}

\title{Effective action, magnetic excitations and quantum fluctuations in 
 lightly doped single layer cuprates}

\author{Alexander I. Milstein}
\affiliation{Budker Institute of Nuclear Physics, 630090 Novosibirsk, Russia}
\author{Oleg P. Sushkov}
\affiliation{School of Physics, University of New South Wales, Sydney 2052, Australia}

\date{\today}

\begin{abstract}

We consider the extended 2D $t-t'-t''-J$ model at zero temperature.
Parameters of the model corresponds to doping by holes.
Using the low doping effective action we demonstrate that the system can
1) preserve the long range collinear antiferromagnetic order,
2) lead to a spin spiral state (static or dynamic),
3) lead to the phase separation instability.
We show that at parameters of the effective action corresponding to the 
single layer cuprate La$_{2-x}$Sr$_x$CuO$_4$
the spin spiral ground state is  realized.
We derive properties of magnetic excitations and calculate quantum fluctuations.
Quantum fluctuations destroy the static spin spiral at  the
critical doping $x_c\approx 0.11$.  This is 
the point of  the quantum phase transition to the spin-liquid state 
(dynamic spin spiral).
The state is still double degenerate with respect to the direction of the
dynamic spiral, so this is a ``directional nematic''.
The superconducting pairing exists throughout the phase diagram and is not
sensitive to the quantum phase transition.
We also compare the calculated neutron scattering spectra with experimental data.

\end{abstract}


\maketitle

\section{Introduction}
The phase diagram of the prototypical cuprate superconductor La$_{2-x}$Sr$_x$CuO$_4$  (LSCO) shows that the magnetic state changes tremendously with Sr doping. The three-dimensional antiferromagnetic (AF) N\'eel order identified~\cite{keimer92} below $325\ \text{K}$ in the parent compound disappears at doping $x\approx0.02$ and gives way to the so-called spin-glass phase which extends up to $x\approx 0.055$. In both, the N\'eel and the spin-glass phase, the system essentially behaves as an Anderson insulator and exhibits only hopping conductivity. Superconductivity then sets in for doping $x\gtrsim 0.055$, see Ref.~\cite{kastner98}.  One of the most intriguing properties of LSCO is the static incommensurate magnetic ordering observed at low temperature in \emph{elastic} neutron scattering  experiments. This ordering manifests itself as a scattering peak shifted with respect to the  antiferromagnetic position. Very importantly, the incommensurate ordering is a generic feature of LSCO. According to experiments in the N\'eel phase, the incommensurability is almost doping independent and  directed along the orthorhombic $b$ axis~\cite{matsuda02}. In the spin-glass phase, the shift is also directed along the $b$ axis, but scales linearly with doping~\cite{wakimoto99,matsuda00,fujita02}. Finally, in the underdoped superconducting region ($0.055 \lesssim x \lesssim 0.12$), the shift still scales linearly with doping, but it is directed along the crystal axes of the tetragonal lattice~\cite{yamada98}. 
Very recent studies reveal also the evolution of inelastic neutron spectra
with doping~\cite{yamada07}.

Near $x=0.12$ certain La-based materials develop a strongly enhanced static
incommensurate magnetic order accompanied by small lattice deformation at the
second order harmonics~\cite{Tranquada95,Tranquada96,Fujita04}, see also
Ref.~\cite{tranquada05} for a review. 


Incommensurate features have also been observed in inelastic neutron 
scattering from  YBa$_2$Cu$_3$O$_{6+y}$ 
(YBCO)~\cite{bourges00,fong00,mook02,hayden04,hinkov04,stock06,hinkov07a}.
In underdoped YBCO there is a rather large uncertainty in the determining of 
the doping level. However, it seems that the incommensurability in YBCO
is about 30-40\% smaller than that in LSCO comparing the same doping level.
In a very recent work~\cite{hinkov07b} the electronic liquid crystal state
in underdoped YBCO has been reported. The state has no static spins, but
nevertheless, it demonstrates a degeneracy with respect to the direction
of the dynamic spin structure.
In addition, there are indications that the electronic liquid crystal state
observed in~\cite{hinkov07b} is very close to a quantum phase transition
to a state with static spins.

The 2D $t-J$ model was suggested two decades ago to describe the essential
low-energy physics of high-$T_{c}$ cuprates~\cite{PWA,Em,ZR}. In its
extended version, this model includes additional hopping matrix elements $%
t^{\prime }$ and $t^{\prime \prime }$ to 2nd and 3rd-nearest Cu
neighbors. The Hamiltonian of the $t-t^{\prime }-t^{\prime \prime }-J$ model
on the square Cu lattice has the form:%
\begin{eqnarray}
H &=&-t\sum_{\langle ij\rangle \sigma }c_{i\sigma }^{\dag }c_{j\sigma
}-t^{\prime }\sum_{\langle ij^{\prime }\rangle \sigma }c_{i\sigma }^{\dag
}c_{j^{\prime }\sigma }-t^{\prime \prime }\sum_{\langle ij^{\prime \prime
}\rangle \sigma }c_{i\sigma }^{\dag }c_{j^{\prime \prime }\sigma }  \notag \\
&+&J\sum_{\langle ij\rangle \sigma }\left( \mathbf{S}_{i}\mathbf{S}_{j}-{%
\frac{1}{4}}N_{i}N_{j}\right) .  \label{H}
\end{eqnarray}%
Here, $c_{i\sigma }^{\dag }$ is the creation operator for an electron with
spin $\sigma $ $(\sigma =\uparrow ,\downarrow )$ at site $i$ of the square
lattice, $\langle ij\rangle $ indicates 1st-, $\langle ij^{\prime }\rangle $
2nd-, and $\langle ij^{\prime \prime }\rangle $ 3rd-nearest neighbor sites. 
The spin operator is $\mathbf{S}_{i}={\frac{1}{2}}c_{i\alpha }^{\dag }%
\mathbf{\sigma }_{\alpha \beta }c_{i\beta }$, and $N_{i}=\sum_{\sigma
}c_{i\sigma }^{\dag }c_{i\sigma }$ with $\langle N_{i}\rangle =1-x$ being
the number density operator. In addition to the Hamiltonian (\ref{H}) there
is the constraint of no double occupancy, which accounts for strong electron
correlations. The values of the parameters of the Hamiltonian (\ref{H}) for
LSCO are known from neutron scattering~\cite{keimer92}, Raman 
spectroscopy~\cite{tokura90}
and ab-initio calculations~\cite{andersen95}.
The values are:
$J \approx 140\,\text{meV}$, $ t\approx 450\,\text{meV}$,  
$t^{\prime } \approx -70\,\text{meV}$ , and
$t^{\prime \prime }\approx 35\,\text{meV}$.
Hereafter we set $J=1$, hence we measure energies in units of $J$.

The idea of spin spirals in the $t-J$ model at finite doping was first suggested in
Ref.~\cite{shraiman88}. The idea had initially attracted a lot of
attention, see e. g. Refs.~\cite{Dom,IF,CM}. 
However, it has been soon realized that there was a fundamental
unresolved theoretical problem of  stability of the spiral~\cite{CM}.
Together with lack of experimental confirmations this was a very
discouraging development.
The observation of static and quasistatic incommensurate peaks in neutron 
scattering caused a renewal of theoretical interest in the idea of spin 
spirals in cuprates~\cite{hasselmann04,sushkov04,juricic04,sushkov05,
lindgard05,juricic06,luscher06,luscher07,luscher07b}. 
It has been realized that in LSCO the charge disorder related to a random 
distribution of Sr ions plays a crucial role
and in the insulating state, $x\leq 0.055$, the disorder qualitatively 
influences the problem of stability of the spiral.
The point is that in the insulating state the mobile holes are not really
mobile, they are trapped in shallow hydrogen-like bound states near Sr ions. 
The trapping leads to the diagonal spin 
spiral~\cite{sushkov05,luscher06,luscher07,luscher07b}.
Percolation of the bound states gives way to superconductivity
and in the percolated state the spin spiral must be directed along crystal axes of
the tetragonal lattice~\cite{sushkov05}. So the percolation concentration is
$x_{per}=0.055$.
The rotation of the direction of the spin spiral is dictated by the 
Pauli exclusion  principle.
The disorder at $x > 0.055$ is still pretty strong. However, unlike in 
the insulating phase,  the disorder does not play a qualitative role and 
therefore in the first approximation  one can disregard it. Thus, we arrive at 
the case of small uniform doping.
This is the problem we address in the present work.

As we already mentioned, the case of an uniform spin spiral (no external disorder)
in a doped quantum antiferromagnet has an inherent theoretical problem.
If considered in the semiclassical approximation, the out-of-plane magnon is 
marginal and in the end this implies an instability of the spin spiral~\cite{CM}.
An attempt to fix the problem by account of quantum fluctuations within the 1/S
spin-wave theory was done in Ref.~\cite{sushkov04}.
We understand now that, while being qualitatively correct, the work~\cite{sushkov04}
did not account for all relevant quantum fluctuations.
The effective action method is much more powerful then the 1/S expansion because
the method accounts for all symmetries exactly and generates a regular expansion in
powers of doping x, this is the true chiral perturbation theory.
This is why in the present work we employ the effective action method.

The structure of the paper is the following. In section \ref{sec:model} we discuss the
effective low-energy action of the modified t-J model.
Section \ref{sec:Neel} addresses the issue of stability of the N\'eel state under doping.
The spiral ground state in the mean-field approximation is considered in section
\ref{sec:Spiral}. The in-plane magnons are discussed in section
\ref{sec:inSpiral} and out-of-plane magnons in section \ref{sec:outSpiral}.
Section \ref{sec:QPT} addresses the quantum fluctuations and the quantum phase
transition to the directional nematic.
Finally discussion and comparison with experiments is presented in the 
section \ref{sec:discussion}.

\section{Effective low-energy action of 2D $t-t'-t''-J$ model at small doping
\label{sec:model}}
 At zero doping (no holes), the $t$-$J$ model is equivalent to the Heisenberg model and describes the Mott insulator 
La$_2$CuO$_4$. The removal of a single electron from this Mott insulator, or in other words the injection of a 
hole, allows the charge carrier to propagate. Single-hole properties of the $t$-$J$ model are well 
understood, see Ref.~\cite{dagotto94} for a review. 
A calculation of the hole dispersion at values of parameters $t$, $t'$, and $t''$
corresponding to the single layer cuprate LSCO has been performed in 
Ref.~\cite{sushkov04} using the Self Consistent Born Approximation (SCBA),
see also Ref.~\cite{luscher06}.
According to this calculation the dispersion of the hole dressed by magnetic 
quantum fluctuations has
minima at the nodal points $\mathbf{q}_{0}=(\pm \pi /2,\pm \pi /2)$, and it is
practically isotropic in the vicinity of each point,
\begin{eqnarray}
\label{eq}
\epsilon \left( \mathbf{p}\right) \approx \frac{1}{2}\beta \mathbf{p}^{2}\ ,
\end{eqnarray}
where ${\bf p}={\bf q}-\mathbf{q}_{0}$.
We set the lattice spacing to unity, 3.81\thinspace \AA $\,\rightarrow $
\thinspace 1.
The SCBA approximation gives $\beta \approx 2.2 \approx 300\,\text{meV}$\ .  
The effective mass corresponding to this value is approximately twice the 
electron mass and this agrees with recent measurement of
Shubnikov - de Haas oscillations~\cite{SdH}.
In the present work we use $\beta$ as a fitting parameter. We will see that to fit
inelastic neutron data at $x=0.1$ we need
\begin{equation}
\label{beta}
\beta\approx 2.7\ .
\end{equation}
This agrees well with the value obtained within the SCBA.
 The quasi-particle residue $Z$  at the minimum of the dispersion is 
$Z\approx0.38$~\cite{sushkov04}. In the full-pocket 
description, where two half-pockets are shifted  by the AF vector ${\bf Q}_{AF}=\left(\pi,\pi\right)$, the 
two minima are located at $S_a=\left(\frac{\pi}{2},\frac{\pi}{2}\right)$ and 
$S_b=\left(\frac{\pi}{2},-\frac{\pi}{2}\right)$. The system is thus somewhat similar to a two-valley semiconductor.

The relevant energy scale for small uniform doping at zero temperature is of the order of 
$xJ \ll J$, relevant momenta are also small, $q \ll 1$.
Hence,  one can simplify the Hamiltonian of the $t$-$J$ model by integrating out 
all high-energy fluctuations. This procedure leads to the effective Lagrangian
or effective action.
The effective Lagrangian has been first discussed quite some time 
ago~\cite{wiegman88,wen89,shraiman88}, see also a recent work~\cite{W}.
That discussion resulted in the kinematic structure of the effective Lagrangian valid
in the static limit~\cite{shraiman88}. This limit is sufficient only for the mean-field 
approximation.
The time-dependent terms that are necessary for excitations and quantum fluctuations 
have been derived only recently~\cite{luscher07b}.
The effective Lagrangian  can be written in terms of 
the bosonic ${\vec n}$-field that describes the staggered component of the copper spins
and in terms of fermionic holons $\psi$. We use the term ``holon'' instead of ``hole'' 
because spin and charge are to some extent separated, see discussion below.
The holon has a pseudospin that originates from two sublattices,
so the fermionic field $\psi$ is a spinor acting on pseudospin.
For the hole-doped case, the effective Lagrangian reads
\begin{eqnarray} 
\label{eq:LL}
{\cal L}&=&\frac{\chi_{\perp}}{2}{\dot{\vec n}}^2-
\frac{\rho_s}{2}\left({\bm \nabla}{\vec n}\right)^2\\
&+&\sum_{\alpha}\left\{ \frac{i}{2}
\left[\psi^{\dag}_{\alpha}{{\cal D}_t \psi}_{\alpha}-
{({\cal D}_t \psi_{\alpha})}^{\dag}\psi_{\alpha}\right]\right.\nonumber\\
&-&\left.\psi^{\dag}_{\alpha}\epsilon_{\alpha}({\bf \cal P})\psi_{\alpha}  
+ \sqrt{2}g (\psi^{\dag}_{\alpha}{\vec \sigma}\psi_{\alpha})
\cdot\left[{\vec n} \times ({\bm e}_{\alpha}\cdot{\bm \nabla}){\vec n}\right]\right\} \ .
\nonumber
\end{eqnarray}
The first two terms in the Lagrangian represent the usual nonlinear $\sigma$ model (NLSM),
the field ${\vec n}$ is the subject of the constraint $n^2=1$.
The magnetic susceptibility and the spin stiffness are
$\chi_{\perp}\approx 0.53/8\approx 0.066$ and $\rho_s \approx 0.18$~\cite{SZ}.
The rest of the Lagrangian in Eq.~(\ref{eq:LL}) represents the fermionic holon field and its
interaction with the ${\vec n}$-field.
The coupling constant is~\cite{IF}, $g\approx Zt\approx 1$.
The index $\alpha=a,b$ (flavor) indicates the  location of the holon in momentum space 
(either in pocket $S_{a}$ or $S_{b}$). 
The kinematic structure of the coupling term was first derived in Ref.~\cite{shraiman88}. 
The operator ${\vec \sigma}$ is a pseudospin that 
originates from the existence of two  sublattices  and 
${\bf e}_{\alpha}=(1/\sqrt{2},\pm 1/\sqrt{2})$ is a unit  vector orthogonal to the face 
of the MBZ where the holon is located. 
Kinetic energy of the holon, $\epsilon_{\alpha}({\bf p})$, is quadratic in the momentum
${\bf p}$ and generally speaking it can be anisotropic.
However, in LSCO the anisotropy is small and we use the isotropic 
approximation (\ref{eq}).

A very important point is that the argument of $\epsilon$ in Eq.~(\ref{eq:LL}) is 
a ``long''
(covariant) momentum~\cite{shraiman88},
\begin{equation}
\label{long1}
{\bf {\cal P}}=-i{\bm \nabla}
+\frac{1}{2}{\vec \sigma}\cdot[{\vec n}\times{\bm \nabla}{\vec n}] \ .
\end{equation}
An even more important point is that the time derivatives that stay in the
kinetic energy of the fermionic field are also ``long'' (covariant)~\cite{luscher07b},
\begin{equation}
\label{long2}
{\cal D}_t=\partial_t
+\frac{i}{2}{\vec \sigma}\cdot[{\vec n}\times{\dot{\vec n}}] \ .
\end{equation}
The covariant time derivatives result in the ``Berry phase term'' ~\cite{luscher07b},
$-\frac{1}{2}\psi_{\alpha}^{\dag}{\vec \sigma}
\psi_{\alpha}\cdot[{\vec n}\times{\dot{\vec n}}]$,
that is crucially important for excitation spectrum and hence for stability of the system
with respect to quantum fluctuations.

Generally speaking, there are also quartic in fermion operators terms in the effective 
Lagrangian.
However, these terms are not important at low doping and therefore we disregard them
in (\ref{eq:LL}).  

The effective Lagrangian (\ref{eq:LL}) is valid regardless if the ${\vec n}$-field is
static or dynamic. In other words it does not matter if the ground state expectation
value of the staggered field is nonzero, $\langle {\vec n}\rangle\ne 0$, or zero,
$\langle {\vec n}\rangle= 0$.
The only condition for validity of (\ref{eq:LL}) is that all dynamic fluctuations
of the ${\vec n}$-field are slow, $1/\tau \ll J$\ , where $\tau$ is the typical
time-scale of the fluctuations.
We will demonstrate below that the dimensionless parameter
\begin{equation}
\label{Omega}
\lambda=\frac{2g^2}{\pi\beta\rho_s}
\end{equation}
plays an important role in the theory.
If $\lambda \leq 1$, the ground state corresponding to the Lagrangian (\ref{eq:LL})
is the collinear N\'eel state and it stays collinear at  any small doping.
If $1\leq \lambda \leq 2$, the N\'eel state is unstable at arbitrary small doping
and the ground state is  static or dynamic spin spiral.
Whether the spin spiral is static or dynamic depends on doping.
If $\lambda \geq 2$, the system is unstable with respect to phase separation and hence
the effective long-wave-length Lagrangian (\ref{eq:LL}) is meaningless. Thus,
\begin{eqnarray}
\label{O12}
&\lambda& \leq 1\ ,  \ \ \ Neel \ \ state\nonumber\\
1\leq &\lambda& \leq 2 \ , \ \ \ Spiral \  state\ , \  static \ or\ dynamic \nonumber\\
&\lambda& \geq 2 , \ \ \ Phase \ \ separation \ .
\end{eqnarray}
For LSCO the value is $\lambda \approx 1.3-1.5$.

We would like to stress once more that spin and charge to some extent are  separated in 
the effective low-energy Lagrangian (\ref{eq:LL}), this is why we use the term ``holon'' 
instead  of ``hole''. 
The holon carries pseudospin, it carries charge, but it does not carry spin in the usual sense.
However, it is not the full spin-charge separation like in 1D models.
To illustrate this point, it is instructive to look at the holon interaction with uniform 
external magnetic field~\cite{luscher06,luscher07b}.
\begin{equation} \label{LLB}
\delta{\cal L}_B=
\frac{1}{2}({\vec B}\cdot{\vec n})
\psi^{\dag}_{\alpha}({\vec \sigma}\cdot{\vec n})\psi_{\alpha} \ .
\end{equation}
Since we only want to stress the spin  dynamics 
this interaction does not include terms that originate from the
long derivative with respect to magnetic vector potential
$-i{\bm \nabla} \to -i{\bm \nabla} -\frac{e}{c}{\bf A}$, describing the interaction of 
the magnetic field with the electric charge.
Clearly the interaction (\ref{LLB}) is quite unusual and this is what we call
``the partial spin-charge separation''. The holon does not interact directly with the 
staggered magnetic field (neutron scattering).

\section{Criterion of stability of the N\'eel phase under doping \label{sec:Neel}}
One can consider the coupling constant $g$ in the Lagrangian (\ref{eq:LL}) as a 
formal parameter.
It is clear that the N\'eel order must be stable
at a sufficiently small $g$,
\begin{equation}
\label{n0n}
{\vec n}\approx {\vec n}_0=(0,0,1)  \ .
\end{equation}
In this case the two hole pockets are populated by holons with pseudospin ``up'' and ``down'',
and hence the Fermi momentum (radius of the pocket) is
\begin{equation}
\label{pfn}
p_F=\sqrt{\pi x} \ ,
\end{equation}
where $x$ is doping.
\begin{figure}[ht]
\includegraphics[width=0.15\textwidth,clip]{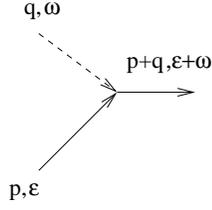}
\caption{Magnon-holon vertex, magnon is shown by the dashed line}
\label{ghh}
\end{figure}
The Lagrangian (\ref{eq:LL}) can be split in the diagonal and offdiagonal part
with respect to transverse spin waves
$n_{\perp}=n_{\pm}=(n_x\pm i n_y)/\sqrt{2}$.
\begin{widetext}
\begin{align} 
\label{LN}
{\cal L}&={\cal L}_0+{\cal L}_{1} \ ,\\
{\cal L}_0&=
\frac{\chi_{\perp}}{2}{\dot{n}_{\perp}}^2-
\frac{\rho_s}{2}\left(1+\frac{\beta x}{4\rho_s}\right)\left({\bm \nabla}{n_{\perp}}\right)^2
+
\sum_{\alpha}\left( \frac{i}{2}
\left[\psi^{\dag}_{\alpha}{\dot \psi}_{\alpha}-
{\dot \psi}_{\alpha}^{\dag}\psi_{\alpha}\right]
-\psi^{\dag}_{\alpha}\epsilon({\bf p})\psi_{\alpha}  
\right) \ ,
\nonumber\\
{\cal L}_{1}&=
\sum_{\alpha}\psi_{\alpha}^{\dag}\left(
-\frac{1}{2}{\vec n}_0[{\dot {\vec n}_{\perp}}\times{\vec \sigma}]
-\frac{\beta}{4}\left\{{{\bm p}},{\vec n}_0[{\bm \partial}{\vec n}_{\perp}\times{\vec \sigma}]\right\}
+\sqrt{2}g{\vec n}_0[({\bm e}_{\alpha}\cdot{\bm \nabla} {\vec n}_{\perp})\times{\vec \sigma}]
\right)\psi_{\alpha} \ . \nonumber
\end{align}
Here $\left\{..., ...\right\}$ stands for the anticommutator.
Using the second quantization representation for the ${\vec n}$-field, 
\begin{equation*} 
n_{\pm}
=\sum_{\bf q}\frac{1}{\sqrt{2\chi_{\perp}\omega_{\bf q}}}\left(
e^{i\omega_{\bf q}t-i{\bf q}\cdot{\bf r}}m_{\pm,\bf q}^{\dag}+
e^{-i\omega_{\bf q}t+i{\bf q}\cdot{\bf r}}m_{\pm,\bf q}\right) \ ,
\end{equation*}
with the magnon creation and annihilation operators 
$m_{\pm,\bf q}^{\dag}$ and $m_{\pm,\bf q}$,
we find the ``bare'' magnon dispersion
\begin{equation}
\label{oq}
\omega_{\bf q}^2=c^2q^2\left(1+\frac{\beta x}{4\rho_s}\right) \ , \ \ \ 
c^2=\frac{\rho_s}{\chi_{\perp}} \ ,
\end{equation}
and the pseudospin flip magnon-holon vertex shown in Fig.~\ref{ghh},  
\begin{equation}
\label{vN}
M=i\sqrt{\frac{2}{\chi_{\perp}}}
\left\{\sqrt{2}g({\bf e}_{\alpha}\cdot{\bf q})+\frac{\omega}{2}+\frac{1}{2}
[\epsilon({\bf p})-\epsilon({\bf p}+{\bf q})]\right\} \ .
\end{equation}
Looking at (\ref{oq}), one can conclude superficially that magnons are
hardened by doping. However, they are not hardened, they are softened.
To see this we need to calculate the magnon polarization operator that is due
to ${\cal L}_1$. The operator reads
\begin{eqnarray}
\label{PN}
{\cal P}_N(\omega,{\bf q})&=&\frac{2}{\chi_{\perp}}\sum_{{\bf p},\alpha}
f_{\bf p}(1-f_{{\bf p}+{\bf q}})
\frac{\left\{\sqrt{2}g({\bf e}_{\alpha}\cdot{\bf q})+\frac{\omega}{2}+\frac{1}{2}
[\epsilon({\bf p})-\epsilon({\bf p}+{\bf q})]\right\}^2}
{\epsilon({\bf p})+\omega-\epsilon({\bf p}+{\bf q})+i0}\nonumber\\
&+&\frac{2}{\chi_{\perp}}\sum_{{\bf p},\alpha}f_{\bf p}(1-f_{{\bf p}-{\bf q}})
\frac{\left\{\sqrt{2}g({\bf e}_{\alpha}\cdot{\bf q})+\frac{\omega}{2}+\frac{1}{2}
[\epsilon({\bf p}-{\bf q})-\epsilon({\bf p})]\right\}^2}
{\epsilon({\bf p})-\omega-\epsilon({\bf p}-{\bf q})+i0}  \ ,
\end{eqnarray}
where $f_{\bf p}$ is the usual Fermi-Dirac step function.
Eq. (\ref{PN}) can be transformed to
\begin{eqnarray}
\label{PN1}
{\cal P}_N(\omega,{\bf q})&=&-\frac{\beta c^2 x}{4\rho_s}q^2+2{\cal P}_0(\omega,{\bf q}) \ ,\nonumber\\
{\cal P}_0(\omega,{\bf q})&=&
\frac{2c^2g^2}{\rho_s}q^2\sum_{{\bf p}}
f_{\bf p}(1-f_{{\bf p}+{\bf q}})
\left(
\frac{1}{\epsilon({\bf p})+\omega-\epsilon({\bf p}+{\bf q})+i0}
+ \frac{1}{\epsilon({\bf p})-\omega-\epsilon({\bf p}+{\bf q})+i0}
\right) \ .
\end{eqnarray}
Then the magnon Green's function reads
\begin{equation} 
\label{GN}
G=\frac{\chi_{\perp}^{-1}}{\omega^2-\omega_{\bf q}^2-{\cal P}(\omega,{\bf q})+i0}
=\frac{\chi_{\perp}^{-1}}{\omega^2-c^2q^2-2{\cal P}_0(\omega,{\bf q})+i0} \ .
\end{equation}
\end{widetext}

The condition of stability of the ground state is the absence of poles of the Green's function
at imaginary $\omega$-axis. Hence this condition is $c^2q^2 \geq -2{\cal P}_0(0,{\bf q})
=\lambda c^2q^2, \ \ at \ \ q\ll p_F$. Doping $x$ does not appear in this criterion.
Thus, as it is stated in (\ref{O12}),
 the  N\'eel state is stable at any small doping if $\lambda \leq 1$.
This criterion has been discussed many times, see e.g.\cite{sushkov04}.
We have rederived it here just to demonstrate how the effective action
technique works in the known situation.

\section{The spiral ground state in the mean-field approximation  \label{sec:Spiral}}
At $\lambda \geq 1$ the minimum energy is realized with the coplanar spiral
\begin{equation}
\label{n0}
{\vec n}_0=(\cos{\bf Q}\cdot{\bf r},\ \sin{\bf Q}\cdot{\bf r},\ 0) \ ,
\end{equation}
where ${\bf Q}\propto (1,0);(0,1)$ is  directed along the CuO bond.
To be specific we assume that ${\bf Q}\propto (1,0)$. 
Due to the holon interaction with the spiral
the holon band is split in two with $\sigma_z=\pm 1$, 
\begin{eqnarray}
\label{epsQ}
\epsilon&\to& -\frac{\Delta}{2}\sigma_z+\frac{1}{2}\beta
\left({\bf p}+\frac{1}{2}{\bf Q}\sigma_z\right)^2 \ ,\nonumber\\
\Delta&=&2gQ \ .
\end{eqnarray}
In the ground state only the  band with $\sigma_z=+1$ is populated. Therefore, the 
Fermi momentum, that is the radius of the Fermi circle in each pocket, is
\begin{equation}
\label{pfs}
p_F=\sqrt{2\pi x} \ .
\end{equation}
The point ${\bf p=0}$ corresponds to ${\bf k}=(\pi/2,\pm\pi/2)$ in the
Brillouin zone. According to (\ref{epsQ}) the center of the filled holon pocket 
($\sigma_z=1)$ is shifted  from this point by  $-\frac{1}{2}{\bf Q}$, and 
the center of the empty pocket ($\sigma_z=-1)$ is shifted by  $\frac{1}{2}{\bf Q}$.
Calculation of energy and its minimization with respect to $Q$ gives
the following value
\begin{equation}
\label{qq}
Q=\frac{g}{\rho_s}x \ .
\end{equation}
The ground state energy of the spiral state is below that of the  N\'eel state only if 
$\lambda \geq 1$.

\section{The in-plane magnons in the spiral state  \label{sec:inSpiral}}
To analyze the stability of the spiral state one needs to go beyond the mean-field
approximation and study excitations and quantum fluctuations in the system.
In this section we consider in-plane magnetic excitations.
An in-plane excitation is described by a small deviation $\varphi=\varphi(t,{\bf r})$ from the uniform spiral ground state~(\ref{n0}),
\begin{equation}
\label{inp}
{\vec n}=(\cos({\bf Q\cdot r}+\varphi),\sin({\bf Q\cdot r}+\varphi),0) \ .
\end{equation}
In the ground state all the holons are in the pseudospin state $\sigma_z=1$.
The in-plane magnons do not change pseudospin, therefore in this section
we set everywhere $\sigma_z=1$.
Substituting expression (\ref{inp}) in the Lagrangian (\ref{eq:LL})
we once more find the diagonal and offdiagonal parts of the Lagrangian
\begin{eqnarray}
\label{LNf}
{\cal L}&=&{\cal L}_0+{\cal L}_{1} \ ,\\
{\cal L}_0&=&
\frac{\chi_{\perp}}{2}{\dot{\varphi}}^2-
\frac{\rho_s}{2}\left(1+\frac{\beta x}{4\rho_s}\right)\left({\bm \nabla}{\varphi}\right)^2
\nonumber\\
&+&\sum_{\alpha}\left( \frac{i}{2}
\left[\psi^{\dag}_{\alpha}{\dot \psi}_{\alpha}-
{\dot \psi}_{\alpha}^{\dag}\psi_{\alpha}\right]
-\psi^{\dag}_{\alpha}\left[
-\frac{\Delta}{2}+\epsilon({\bf l}^2)\right]\psi_{\alpha}  
\right) \ ,
\nonumber\\
{\cal L}_{1}&=&
\sum_{\alpha}\psi_{\alpha}^{\dag}\psi_{\alpha}\left(\sqrt{2}g({\bm e}_{\alpha}\cdot{\bm \nabla})
\varphi-\frac{1}{2}{\dot \varphi}-\frac{\beta}{4}\left\{{\bm l},{\bm \partial}\varphi\right\}
\right) \ .
 \nonumber
\end{eqnarray}
Here ${\bf l}={\bf p}+{\bf Q}/2$ is shifted momentum and
$\left\{..., ...\right\}$ stands for anticommutator.
Thus, the ``bare'' magnon dispersion is given by the same Eq. (\ref{oq}) as for the N\'eel
state, but the magnon-holon vertex is smaller than (\ref{vN}) by the factor $\sqrt{2}$,
\begin{align}
\label{vf}
M&=i\sqrt{\frac{1}{\chi_{\perp}}}
\left\{\sqrt{2}g({\bf e}_{\alpha}\cdot{\bf q})+\frac{\omega}{2}+\frac{1}{2}
[\epsilon({\bf l})-\epsilon({\bf l}+{\bf q})]
\right\} \ .
\end{align}
A calculation similar to that performed in section \ref{sec:Neel} for the N\'eel state
gives the following Green's function for the field $\varphi$ that describes the in-plane magnon
\begin{equation} 
\label{Gin}
G_{in}=\frac{\chi_{\perp}^{-1}}{\omega^2-c^2q^2-{\cal P}_0+i0} \ ,
\end{equation}
where ${\cal P}_0$ is given by Eq. (\ref{PN1}) with the Fermi momentum (\ref{pfs}).
At zero frequency and  at small $q$, $q\ll p_F$, the polarization operator is equal to
${\cal P}_0(0,{\bf q})=-\frac{\lambda}{2} c^2q^2$.
Therefore, the ground state is getting unstable (poles of the Green's function at 
imaginary $\omega$-axis)
at $\lambda \geq 2$. This is the instability with respect to phase 
separation~\cite{CM,sushkov04}
and it is fatal for the effective long-wave-length Lagrangian (\ref{eq:LL}).
Thus, the spiral state is stable at $1\leq \lambda \leq 2$, see Eq. (\ref{O12}).
We also present here an explicit expression for the polarization operator
\begin{widetext}
\begin{eqnarray}
\label{pol0}
{\text {Re}}\ {\cal P}_0(\omega,q)&=&-\frac{c^2g^2}{\pi\beta^2\rho_s}\left\{\beta q^2
-R_1\sqrt{1-R_0^2/R_1^2}\ \ \theta(1-R_0^2/R_1^2)
-R_2\sqrt{1-R_0^2/R_2^2}\ \ \theta(1-R_0^2/R_2^2)\right\}\ , \nonumber\\
{\text {Im}}\ {\cal P}_0(\omega,q)&=&-\frac{c^2g^2}{\pi\beta^2\rho_s}
\left\{ \theta(R_0^2-R_1^2)\ \sqrt{R_0^2-R_1^2}-
\sqrt{R_0^2-R_2^2}\ \theta(R_0^2-R_2^2)\right\}\ , \nonumber\\
R_0&=&\beta q p_F \ , \ \ \ R_1=\frac{1}{2}\beta q^2-\omega \ , \ \ \ 
R_2=\frac{1}{2}\beta q^2+\omega \ .
\end{eqnarray}
\end{widetext}
The Fermi momentum $p_F$ is given by (\ref{pfs}), and $\theta(x)$ is the usual step function.
\begin{figure}[ht]
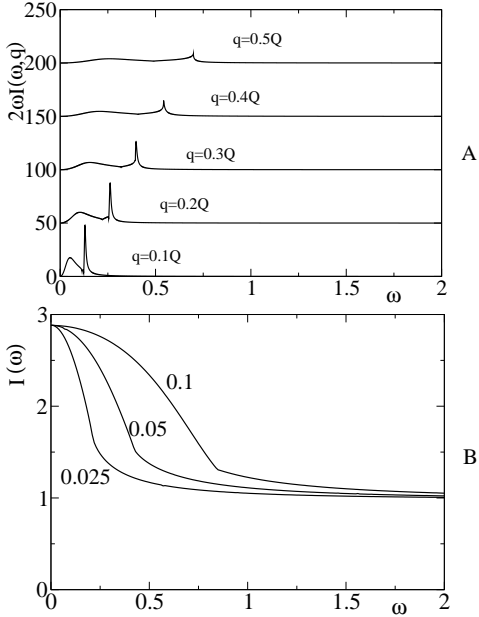

\includegraphics[width=0.35\textwidth,clip]{Iin1.eps}
\includegraphics[width=0.35\textwidth,clip]{Iin2.eps}
\caption{{\bf A}: Plots of $2\omega I_{in}(\omega,q)$
versus energy for different values of momentum $q$ (offsets).
$I_{in}(\omega,q)$ is the in-plane magnon spectral 
density (\ref{ioq}). The plots are presented for doping $x=0.1$, and
$\beta=2.7$, $g=1$. Values of $q$ are given in units of the incommensurate vector
$Q$, see (\ref{qq}).\\
{\bf B}: $q$-integrated in-plane spectral density (\ref{io}) for dopings 
$x=0.025$, $0.05$, and $0.1$. The parameters of the effective Lagrangian are
 $\beta=2.7$, $g=1$.
}
\label{gfi}
\end{figure}

It is convenient to define the magnon spectral density as 
\begin{align}
\label{ioq}
I_{in}(\omega,{\bf q})=-4\rho_s \text{Im} G_{in}(\omega,q) \ .
\end{align}
Plots of $2\omega I_{in}(\omega,{\bf q})$ versus $\omega$ are presented
in Fig.~\ref{gfi}A for different values of momentum $q$ (offsets).
The doping is $x=0.1$, and $\beta=2.7$, $g=1$.
The narrow peak is the $\delta$-function broadened ``by hands'' to fit in the picture
size.
The corresponding quasiparticle residue is rather small, say for $q=0.1Q$ in Fig.~\ref{gfi}A
the residue is $Z=0.39$ and it very quickly dies out at larger values of $q$. The magnon
``dissolves'' in the particle-hole continuum.

The $q$-integrated in-plane magnon spectral density
\begin{equation}
\label{io}
I_{in}(\omega)=\int I_{in}(\omega,{\bf q}) \frac{d^2 q}{(2\pi)^2}
\end{equation}
is plotted in Fig.~\ref{gfi}B for doping $x=0.025$, $x=0.05$, and $x=0.1$.
For zero energy the value of the $q$-integrated spectral density is independent
of doping and equals to $I_{in}(0)=1/(1-\lambda/2)$.
\begin{figure}[ht]
\includegraphics[width=0.35\textwidth,clip]{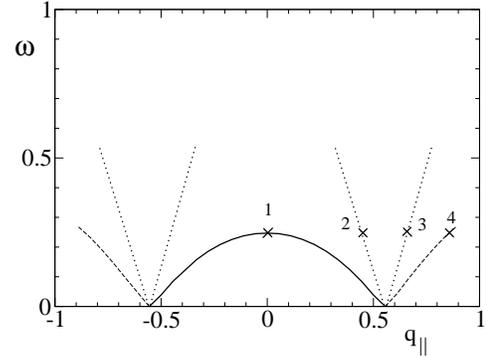}
\caption{The magnon dispersion along {\bf Q}. The parameters are $x=0.1$, $\beta=2.7$,
$g=1$.
The out-of-plane excitation for $|q|\leq Q$ is shown by the solid line and the 
out-of-plane excitation for $|q|\geq Q$ is shown by the dashed line.
The in-plane excitation is shown by the dotted line.
The quasiparticle residue decays very quickly outside of the dome shown by the solid line.
The quasiparticle residue at point 1 at the top of the dome is $Z=0.8$
while the quasiparticle residue at point 4 that is outside of the dome
at the same height is just $Z=0.13$.
The quasiparticle residue of the in-plane magnon at the same frequency as
the dome height (points 2 and 3) is $Z=0.15$.
}
\label{spe}
\end{figure}

To calculate the in-plane spectral density that can be observed in neutron scattering 
one needs to shift momenta.
The Hamiltonian describing the interaction of the neutron spin ${\vec S}^N$ with the 
${\vec n}$-field reads
\begin{equation} \label{n}
H^N\propto {\vec S}^N\cdot{\vec n}=S^N_z n_z+\frac{1}{2}\left(S^N_+n_-+S^N_-n_+\right) \ .
\end{equation}
After the substitution of the in-plane excitation~(\ref{inp}), the above Hamiltonian reads
\begin{align*} 
H^N&\propto\frac{1}{2}S^N_+e^{-i({\bf Q}\cdot{\bf r}+\varphi)}
+\frac{1}{2}S^N_-e^{i({\bf Q}\cdot{\bf r}+\varphi)} \nonumber \\
&\to\frac{1}{2}e^{i{\bf k}\cdot{\bf r}}\left\{S^N_+e^{-i{\bf Q}\cdot{\bf r}}(1-i\varphi)
+S^N_-e^{i{\bf Q}\cdot{\bf r}}(1+i\varphi)\right\} \ ,
\end{align*}
where ${\bf k}$ is the momentum transfer and ${\bf Q}$ the momentum shift due to the spiral ground state. The scattering probability for unpolarized neutrons is given by
\begin{eqnarray} 
\label{i1}
{\cal I}_{in}(\omega,{\bf k})=\frac{1}{2}\left[I_{in}(\omega,{\bf k}-{\bf Q})
+I_{in}(\omega,{\bf k}+{\bf Q})\right] \ .
\end{eqnarray}
In Fig.~\ref{spe} we show by dotted lines the brunches of linear dispersion
that correspond to the quasiparticle peak in the  spectral  function 
$I_{in}(\omega,{\bf q})$ plotted  in Fig.~\ref{gfi}A. 
The dispersion is very steep, steeper than the bare magnon velocity $c$, and the 
corresponding intensities are very low.

\section{The out-of-plane magnons  in the spiral state  \label{sec:outSpiral}}
Dynamics of out-of-plane magnons are the most complicated ones. 
Stability of the spiral state
was questioned because of the ``marginal'' character of the out-of-plane 
excitations if considered in semiclassical 1/S-approximation~\cite{shraiman88,CM}.
The effective action technique allows us to resolve the problem because the technique
accounts exactly all the symmetries.
\begin{figure}[ht]
\includegraphics[width=0.35\textwidth,clip]{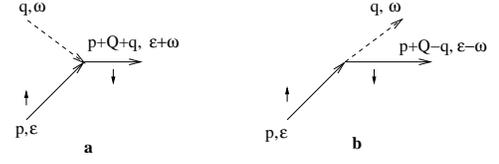}
\caption{Magon-holon vertexes with pseudospin flip, magnon is shown by the dashed line}
\label{vertexflip}
\end{figure}
For the out-of-plane excitation  let us write the ${\vec n}$-field as 
\begin{equation*} 
{\vec n}=(\sqrt{1-n_z^2}\cos{\bf Q}\cdot{\bf r}, \sqrt{1-n_z^2}\sin{\bf Q}\cdot{\bf r}, n_z)\ ,
\end{equation*}
and substitute this expression into the effective Lagrangian~(\ref{eq:LL}).
Neglecting cubic and higher order terms in $n_z$, we get the diagonal and the
offdiagonal parts of the Lagrangian
\begin{widetext}
\begin{align}
\label{LNz}
{\cal L}&={\cal L}_0+{\cal L}_{1} \ ,\\
{\cal L}_0&=
\frac{\chi_{\perp}}{2}{\dot{n}_z}^2-
\frac{\rho_s}{2}\left(1+\frac{\beta x}{4\rho_s}\right)
\left[Q^2n_z^2+\left({\bm \nabla}{n}_z\right)^2\right]
+\sum_{\alpha}\left( \frac{i}{2}
\left[\psi^{\dag}_{\alpha}{\dot \psi}_{\alpha}-
{\dot \psi}_{\alpha}^{\dag}\psi_{\alpha}\right]
-\psi^{\dag}_{\alpha}\left[
-\frac{\Delta}{2}\sigma_z+\frac{\beta}{2}\left({\bf p}+\frac{1}{2}{\bf Q}\sigma_z\right)^2
\right]\psi_{\alpha}  
\right) \ ,
\nonumber\\
{\cal L}_{1}&= -\sum_{\alpha} \psi^{\dag}_{\alpha}
 \frac{\sigma_+}{2}\left(e^{-i{\bf Q}\cdot{\bf r}}\left[g[Qn_z-i\sqrt{2}({\bm e}_{\alpha}
\cdot {\bm \nabla})n_z]+\frac{i}{2}{\dot  n}_z\right]
-\frac{\beta}{4}\left\{{\bm p},e^{-i{\bf Q}\cdot{\bf r}}[{\bm Q}n_z-i{\bm \partial}n_z]\right\}
\right)\psi_{\alpha} +h.c. \ .\nonumber
\end{align}
Here $\sigma_+=\sigma_x+i\sigma_y$ and the bracket $\left\{..., ...\right\}$ stands 
for anticommutator.
According to (\ref{LNz}), the ``bare'' magnon dispersion in this case is
\begin{equation}
\label{oqz}
\omega_{b,\bf q}^2=c^2(Q^2+q^2)\left(1+\frac{\beta x}{4\rho_s}\right) \ .
\end{equation}
The interaction ${\cal L}_{1}$ generates the following two pseudospin-flip vertexes shown in Fig.~\ref{vertexflip} \ ,
\begin{eqnarray}
\label{vz}
M_a&=&i\sqrt{\frac{1}{\chi_{\perp}}}
\left\{g[Q-\sqrt{2}({\bf e}_{\alpha}\cdot{\bf q})]-\frac{\omega}{2}
-\frac{\beta}{4}
\left[(2{\bf p}+{\bf Q}+{\bf q})\cdot({\bf Q}-{\bf q})\right]\right\} \ ,\nonumber\\
M_b&=&i\sqrt{\frac{1}{\chi_{\perp}}}
\left\{g[Q+\sqrt{2}({\bf e}_{\alpha}\cdot{\bf q})]+\frac{\omega}{2}
-\frac{\beta}{4}
\left[(2{\bf p}+{\bf Q}-{\bf q})\cdot({\bf Q}+{\bf q})\right]\right\} \ .
\end{eqnarray}
Hence, the magnon polarization operator  determined by the vertexes reads
\begin{eqnarray}
\label{PZ}
&&{\cal P}(\omega,{\bf q})=\\
&&=\frac{2}{\chi_{\perp}}\sum_{{\bf l}}
f_{\bf l}
\left(\frac{\left[g(Q-q_{||})-\frac{\omega}{2}-\frac{\beta}{4}
(2{\bf l}+{\bf q})\cdot({\bf Q}-{\bf q})\right]^2+g^2q_{\perp}^2}
{\epsilon({\bf l})-\epsilon({\bf l}+{\bf q})+\omega-\Delta+i0}
+\frac{\left[g(Q+q_{||})+\frac{\omega}{2}-\frac{\beta}{4}
(2{\bf l}-{\bf q})\cdot({\bf Q}+{\bf q})\right]^2+g^2q_{\perp}^2}
{\epsilon({\bf l})-\epsilon({\bf l}-{\bf q})-\omega-\Delta+i0}\right) \ .
\nonumber
\end{eqnarray}
Here $q_{||}$ and $q_{\perp}$ are components of momentum parallel and perpendicular
to ${\bf Q}$, respectively; $f_{\bf l}$ is the Fermi-Dirac step function and
${\bf l}={\bf p}+{\bf Q}/2$ is the shifted momentum.
Eq. (\ref{PZ}) can be transformed to
\begin{eqnarray}
\label{PZ1}
&&{\cal P}(\omega,{\bf q})=-\frac{\beta c^2 x}{4\rho_s}q^2
-c^2Q^2+\\
&+&\frac{2c^2}{\rho_s}\sum_{{\bf l}}
f_{\bf l}
\left(\left[gq_{||}+\frac{\beta}{2}{\bf Q}\cdot({\bf l}+{\bf q}/2)\right]^2
+g^2q_{\perp}^2\right)
\left(
\frac{1}{\epsilon({\bf l})-\omega-\epsilon({\bf l}+{\bf q})-\Delta+i0}
+\frac{1}{\epsilon({\bf l})+\omega-\epsilon({\bf l}+{\bf q})-\Delta+i0}
\right) \ .\nonumber
\end{eqnarray}
This form is explicitly symmetric with respect to $\omega \to -\omega$ and
${\bf q}\to -{\bf q}$.
Integration in (\ref{PZ1}) leads to the following magnon Green's function
\begin{eqnarray} 
\label{GZ}
G_{out}&=&\frac{\chi_{\perp}^{-1}}{\omega^2-\omega_{b,\bf q}^2-{\cal P}(\omega,{\bf q})+i0}\\
&=&\chi_{\perp}^{-1}\left[\omega^2-2c^2Q^2\frac{q_{\perp}^2}{q^2}
\left(1-\frac{Q^2}{q^2}\right)
-c^2q^2\left(1-\frac{Q^2}{q^2}\right)^2+\frac{c^2}{\pi\beta^2\rho_s}(F_++F_-)
+i0\right]^{-1}\,,\nonumber
\end{eqnarray}
where
\begin{eqnarray}
&&\mbox{Re}\,F_+=
\frac{A}{4q^2}R_{\Delta}
\left[1-\sqrt{1-\frac{R_0^2}{R_\Delta^2}}\,\theta
\left(1-\frac{R_0^2}{R_\Delta^2}\right)\right]
+\frac{Q^2q_\perp^2}{6q^6}R_\Delta^3
\left[1-\sqrt{1-\frac{R_0^2}{R_\Delta^2}}\left(1+\frac{R_0^2}{2R_\Delta^2}\right)
\,\theta\left(1-\frac{R_0^2}{R_\Delta^2}\right)\right]\,,\nonumber\\
&&\mbox{Im}\,F_+=
\sqrt{R_0^2-R_\Delta^2}\,\theta
\left(1-\frac{R_\Delta^2}{R_0^2}\right)
\left\{\frac{A}{4q^2}
+\frac{Q^2q_\perp^2}{6q^6}
\left(\frac{R_0^2}{2}+R_\Delta^2\right)\right\}\,,\nonumber\\
&&A=4g^2q^2+q_\parallel^2Q^2\left[\frac{\beta^2}{4}+\frac{2g\beta}{Q}-
\frac{R_\Delta}{q^2}\left(\frac{4g}{Q}+\beta\right)\right]+
\frac{R_\Delta^2Q^2}{q^4}(q_\parallel^2-q_\perp^2)\, ,\nonumber\\
&&R_\Delta =\Delta-\omega+\frac{1}{2}\beta q^2\, .
\end{eqnarray}
\end{widetext}
Here $\theta(x)$ is the step function and $R_0$ is defined in (\ref{pol0}).
 The function $F_-$ is obtained from
$F_+$ by the replacement $\omega \to -\omega$ in $R_{\Delta}$.

\begin{figure}[ht]
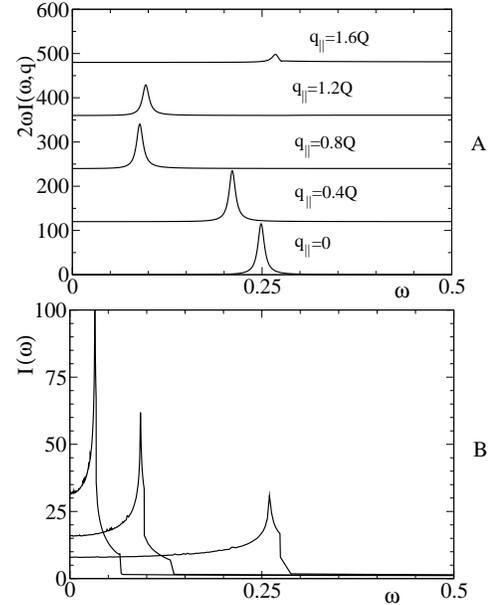

\includegraphics[width=0.35\textwidth,clip]{Iout1.eps}
\hspace{10pt}
\includegraphics[width=0.35\textwidth,clip]{Iout2.eps}
\caption{
{\bf A}: Plots of $2\omega I_{out}(\omega,q)$
versus energy for different values of momentum $q$ (offsets).
$I_{out}(\omega,q)$ is the out-of-plane magnon spectral 
density (\ref{ioq1}). The plots are presented for doping $x=0.1$ and
$\beta=2.7$, $g=1$. Values of $q$ are given in units of the incommensurate vector
$Q$, see (\ref{qq}).\\
{\bf B}: $q$-integrated in-plane spectral density (\ref{iot}) for dopings 
$x=0.025$, $0.05$, and $0.1$. The parameters are $\beta=2.7$, $g=1$.
}
\label{fig:outofplane}
\end{figure}

\begin{figure}[ht]
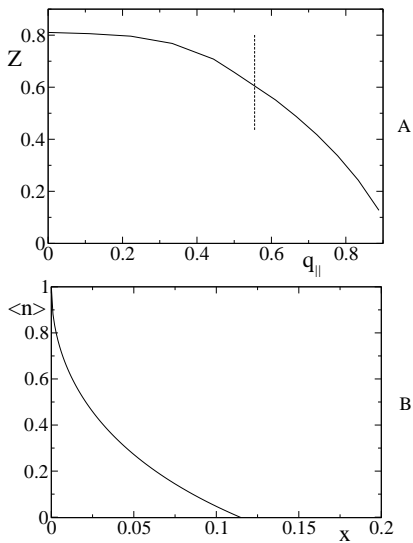

\includegraphics[width=0.3\textwidth,clip]{Zdata1.eps}
\includegraphics[width=0.3\textwidth,clip]{Stagg.eps}
\caption{{\bf A}: The quasiparticle residue versus momentum for the out-of-plane 
magnon for the direction along the spiral, $q=q_{||}$, $q_{\perp}=0$. 
The vertical line shows the momentum where the dispersion vanishes.
The doping is $x=0.1$.\\
{\bf B}: The static component of $n$-field versus doping.
The parameters are $g=1$, $\beta=2.7$.}
\label{Zout}
\end{figure}

We define the out-of-plane magnon spectral density as 
\begin{align}
\label{ioq1}
I_{out}(\omega,{\bf q})=-4\rho_s \text{Im} \ G_{out}(\omega,q) \ .
\end{align}
Plots of $2\omega I_{out}(\omega,{\bf q})$ versus $\omega$ are presented
in Fig.~\ref{fig:outofplane}A for different values of momentum $q_{||}$ (offsets)
and $q_{\perp}=0$. The doping is $x=0.1$.
The narrow peak is the $\delta$-function broadened ``by hands'', the effective width is the same as that for in-plane magnons in Fig.~\ref{gfi}.
The corresponding quasiparticle dispersion $\Omega_{\bf q}$ is plotted in 
 Fig.~\ref{spe} for direction along 
${\bf Q}$, (i.e. $q=q_{||}$, $q_{\perp}=0$) and for $x=0.1$.
The part for $|q|\leq Q$ is shown by the solid line and the part for
$|q|\geq Q$ is shown by the dashed line. We do it to stress that the 
quasiparticle residue decays very quickly outside of the dome.
Plot of the residue is shown in Fig.~\ref{Zout}A.
To illustrate intensities we compare points 1-4 in  Fig.~\ref{spe} 
which correspond to different brunches of dispersion with the same frequency.
The quasiparticle residue at point 1 at the top of the dome is $Z=0.8$
while the quasiparticle residue at point 4 that is outside of the dome
at the same height is $Z=0.13$.
The quasiparticle residue of the in-plane magnon at the same frequency as
the dome height (points 2 and 3 in Fig.~\ref{spe}) is $Z=0.15$.

An analysis of Eq.~(\ref{GZ}) gives the following approximate formulas for the 
dispersion of the out-of-plane magnon and for the corresponding quasiparticle residue
\begin{eqnarray}
\label{OZ}
\Omega_{\bf q}^2&\approx&
\frac{\beta x Q^2c^2}{4\rho_s}\left(1-\frac{1}{\lambda}\right)
\frac{\left(1-\frac{q^2}{Q^2}\right)^2+2\frac{q_\perp^2}{Q^2}}
{1+\frac{c^2q^2}{4g^2Q^2}} \ , \nonumber\\
Z&\approx&\frac{1}{1+\frac{c^2q^2}{4g^2Q^2}} \ .
\end{eqnarray}
These formulas have very limited region of validity since, as we already pointed out,
at larger $q$ the magnon dissolves in the particle-hole continuum.
At $x=0.1$ the equation (\ref{OZ}) for $\Omega_{\bf q}$ agrees reasonably
well with the result of numerical calculation shown in Fig.~\ref{spe}.
At the same time the formula (\ref{OZ}) for the quasiparticle residue
only poorly agrees with numerics shown in Fig.~\ref{Zout}A.
Certainly at very small doping Eq. (\ref{OZ}) is accurate.

The $q$-integrated out-of-plane magnon spectral density
\begin{equation}
\label{iot}
I_{out}(\omega)=\int I_{out}(\omega,{\bf q}) \frac{d^2 q}{(2\pi)^2}
\end{equation}
is plotted in Fig.~\ref{fig:outofplane}B for doping 
$x=0.025$, $x=0.05$, and $x=0.1$.
It is peaked at energy $E_{cross}$ corresponding to the top of the dome
in  Fig.~\ref{spe}. 
Interestingly, the spectral density decays almost abruptly to its high 
frequency asymptotic value $I(\omega)\to 1$ as soon as the magnon
is dissolved in the particle-hole continuum.
The decay of the in-plane $q$-integrated spectral density shown in
Fig.~\ref{gfi} is not that steep.

\section{Quantum fluctuations and quantum phase transition to the
dynamic spiral phase (directional nematic)}\label{sec:QPT}
Due to in-plane and out-of plane quantum fluctuations the static component of the 
staggered field  ${\vec n}$ is reduced,
\begin{equation}
\label{nred}
\langle n \rangle \approx 1-\frac{1}{2}\langle \varphi^2\rangle
-\frac{1}{2}\langle n_z^2\rangle \ .
\end{equation}
Expectation values $\langle \varphi^2\rangle$ and
$\langle n_z^2\rangle$ can be expressed in terms of Green's function
or in terms of q-integrated spectral densities
\begin{eqnarray}
\label{fin}
\langle \varphi^2\rangle&=&-\sum_{\bf q}\int\frac{d\omega}{2\pi i}
G_{in}(\omega,{\bf q})=\frac{1}{4\rho_s}\int\frac{d\omega}{2\pi}
I_{in}(\omega) , \\
\langle n_z^2\rangle&=&-\sum_{\bf q}\int\frac{d\omega}{2\pi i}
G_{out}(\omega,{\bf q})=\frac{1}{4\rho_s}\int\frac{d\omega}{2\pi}
I_{out}(\omega) \ .\nonumber
\end{eqnarray}
These expressions must be renormalized by subtraction of the 
ultraviolet-divergent contribution that corresponds to the undoped $\sigma$-model.
The physical meaning of relations (\ref{nred}) and (\ref{fin}) is very simple: 
the reduction
of static response is transferred to the dynamic response.
The most important contribution to quantum fluctuations comes from
out-of-plane excitations with momenta $q\sim Q \propto x$.
To find this contribution we use the Green's function 
$G_{out}\approx\frac{Z_{\bf q}}{\omega^2-\Omega_{\bf q}^2}$, where
$Z$ and $\Omega$ are given by Eq. (\ref{OZ}).
This gives
\begin{eqnarray}
\label{nz2}
&&<n_z^2>\to\frac{gc}{\pi^2\sqrt{\beta }\rho_s^{3/2}
\sqrt{1-1/\lambda}}{\cal B}\sqrt{x} \ , \\
&&{\cal B}=\frac{1}{2}\int_0^\infty\int_0^{\pi/2}\frac{dt d\varphi}
{\sqrt{\left(1+\frac{c^2}{4g^2}t\right)\left[(1-t)^2+2t\cos^2\phi\right]}}\, .
\nonumber
\end{eqnarray}
Thus, the leading term in the quantum fluctuation scales as $\propto \sqrt{x}$.
The subleading contribution to the quantum fluctuation  scales is $x$.
To find it we have performed numerical integration in Eq. (\ref{fin}) 
using q-integrated spectral densities $I_{out}$ and $I_{in}$ calculated in 
sections \ref{sec:inSpiral} and \ref{sec:outSpiral}, see Fig.~\ref{gfi}B and 
Fig.~\ref{fig:outofplane}B. This gives
\begin{equation}
\label{nred1}
\langle n \rangle \approx 1
-\frac{gc}{2\pi^2\sqrt{\beta }\rho_s^{3/2}\sqrt{1-1/\lambda}}{\cal B}\sqrt{x}
+2.6x \ .
\end{equation}
Certainly the coefficient in the subleading x-term depends on parameters 
(a rather weak dependence). The value 2.6 in (\ref{nred1}) corresponds to 
$g=1$ and $\beta=2.7$. The plot of $\langle n \rangle$ versus doping $x$ at 
these values of parameters  is presented in Fig.~\ref{Zout}B.

According to Fig.~\ref{Zout}B the static component of ${\vec n}$
vanishes at $x=x_c\approx0.11$. This is a quantum critical point for
transition to the dynamic spiral. In this phase there isn't
a spontaneous direction of the ${\vec n}$-field, $\langle {\vec n}\rangle=0$,
but the spiral direction (1,0) or (0,1) is still spontaneously
selected. In our opinion, this is the ``nematic phase'' observed
in Ref.~\cite{hinkov07b}.
Clearly the value $x_c\approx 0.11$ is an approximate value.
In doing the spin-wave theory we assume that $\langle \varphi^2\rangle,
\langle n_z^2\rangle \ll 1$, but then, to find the critical point we
extend this consideration to $\langle n_z^2\rangle \sim 1$.
This extension brings some uncertainty in the value of $x_c$.
We also would like to note that the value of $x_c$ is rather sensitive
to parameters. The main sensitivity comes from $\sqrt{1-1/\lambda}$
in the denominator in Eq. (\ref{nred1}). 
The value of $\lambda$ given by Eq.~(\ref{Omega}) is closely related to the
value of the incommensurate vector $Q$ given by Eq.~(\ref{qq}).
Our estimate of $x_c$ is valid for LSCO.

At $x \ge x_c$ the spin-wave pseudogap is opened. To describe the gapped
phase we use the Takahashi approach~\cite{takahashi}, see also 
Ref.~\cite{chandra}.
Idea of this approach is to impose constraint $\langle n\rangle=0$
using the Lagrange multiplier method.
So we introduce an additional term in the effective Lagrangian
\begin{equation}
\label{Lmult}
\delta {\cal L}=\chi_{\perp}\Delta_s^2\left(1-\frac{1}{2}\varphi^2
-\frac{1}{2}n_z^2\right)\ ,
\end{equation}
where $\Delta_s$ is technically the Lagrange multiplier, and physically
this is the spin-wave pseudogap.
The value of $\Delta_s$ must be determined from the condition 
\begin{equation}
\label{n00}
\langle n \rangle=1-\frac{1}{2}\langle \varphi^2\rangle
-\frac{1}{2}\langle n_z^2\rangle=0\ .
\end{equation}
The in-plane quantum fluctuation $\langle\varphi^2\rangle$ is only
very weakly (quadratically) dependent on the pseudogap $\Delta_s$.
The out-of-plane fluctuation $\langle n_z^2\rangle$ contains a term
that depends on $\Delta_s$ linearly. The term comes from the
$\sqrt{x}$-contribution in  Eqs.  (\ref{nz2}),(\ref{nred1}).
\begin{figure}[ht]
\includegraphics[width=0.35\textwidth,clip]{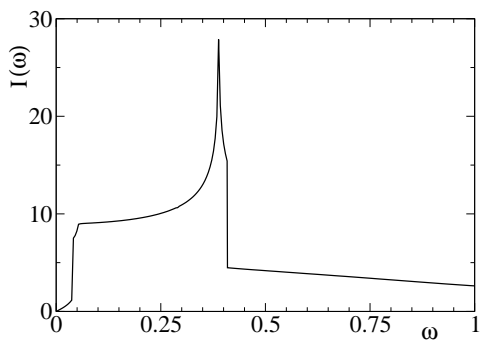}
\caption{The $q$-integrated  magnon spectral density $I=I_{in}+I_{out}$  in the gapped
``nematic'' phase for $x=0.13$ and  $x_c=0.11$.}
\label{Inematic}
\end{figure}
To account for the pseudogap one needs to replace the expression
in square brackets under the square root in ${\cal B}$, see Eq. (\ref{nz2}), by
\begin{equation}
\left[(1-t)^2+2t\cos^2\phi\right]
+\frac{4g}{c^2\beta(1-1/\lambda)Q^3}\Delta_s^2 \ .
\end{equation}
A simple calculation with parameters $g=1$ and $\beta=2.7$ shows that
the condition (\ref{n00}) results in the following pseud0gap
\begin{equation}
\label{pseudo}
\Delta_s \approx 2.5(x-x_c) \ .
\end{equation}
This formula is valid only very close to the critical point.
In this problem one cannot expect a high accuracy from the Takahashi-like approach.
Therefore,  the slope 2.5 in Eq.~(\ref{pseudo}) is rather approximate.
Finally, in Fig.~\ref{Inematic} we present the plot of the q-integrated magnon
spectral density $I_{in}(\omega)+I_{out}(\omega)$
for $x=0.13$ and  $x_c=0.11$.
The figure clearly demonstrates that $\Delta_s$ is a pseudogap since
there is some spectral weight at $\omega \leq \Delta_s$.

\section{Discussion and comparison with experiment \label{sec:discussion}}

There are several points that can be directly compared with experiment.
The incommensurability vector $Q$ is given by equation (\ref{qq}).
It depends on the coupling constant $g$. Fit of experimental
incommensurability~\cite{yamada98} gives $g\approx 1$ and this agrees remarkably
well with prediction of the $t-t'-t''-J$ model.

An important dynamical parameter is $E_{cross}$ which is the height of the dome 
in Fig.~\ref{spe}. This parameter has been systematically studied
very recently in inelastic neutron scattering~\cite{yamada07}.
The experimental values are presented in Table I.
\begin{table}[ht]
\begin{tabular}{|c|ccc|cc|}
\hline
& & & & &\\ 
$ x$ & 0.025 & 0.04 & 0.05 & 0.07& 0.1\\
& & & & &\\ 
$E_{cross}(\mbox{meV}) $ & \ \ 7 $^{+4}_{-2}$\ \ & 
\ \ 15 $^{+7}_{-3}$\ \ & \ \ 20 $^{+6}_{-5}$\ \ 
&\ \ 23 $^{+9}_{-7}$\ \ &\ \ 40 $^{+5}_{-5}$\ \ \\ 
& & & & &\\ 
\hline
$\mbox{phase}$&
\multicolumn{3}{c|} {\mbox{insulator}}&
\multicolumn{2}{c|}{\mbox{superconductor}}\\
$\mbox{spiral direction}$&
\multicolumn{3}{c|} {\mbox{diagonal}}&
\multicolumn{2}{c|}{\mbox{parallel}}\\
$\mbox{theory}$&
\multicolumn{3}{c|} {\mbox{Refs.~\cite{luscher07,luscher07b}}}&
\multicolumn{2}{c|}{\mbox{present work}}\\
\hline
\end{tabular}\\
\caption{\it LSCO: Experimental values~\cite{yamada07}
 of $E_{cross}$ versus doping $x$.}
\end{table}
As soon as the coupling constant $g$ is found from the experimental 
incommensurability $Q$ we can fit $E_{cross}$.
As we already pointed out above, the present theory is applicable to LSCO
at $x \geq x_{per}\approx 0.055$. According to Eq.~(\ref{OZ})
\begin{equation}
\label{ecross}
E_{cross}\approx 
\sqrt{\frac{\beta x Q^2c^2}{4\rho_s}\left(1-\frac{1}{\lambda}\right)}
\ .
\end{equation}
Comparing this formula with data at  $x=0.07$ and $x=0.1$ in Table~I  we find that 
$\beta=2.65(1\pm 0.1)$.
 This value agrees reasonably well with the value 
$\beta\approx 2.2$ that follows from the $t-t'-t''-J$ model.
Note that in principle the inverse mass $\beta$ can be somewhat dependent on doping. However, the data with error bars are quite consistent with x-independent 
$\beta$.

Let us discuss also the data at $0.02 \leq x \leq x_{per}=0.055$
that is relevant to the insulating phase with diagonal disordered spin spiral.
The corresponding theory has been developed in Refs.~\cite{luscher07,luscher07b}.
The incommensurability in this case is $Q=\sqrt{2}gx/\rho_s$. To fit the
experimental incommensurability we need $g\approx 0.7$.
This is somewhat smaller than the value in the conducting phase.
We believe that the reduction of $g$ is due to interaction with phonons.
The point is that $g=Zt$, where $Z$ is the quasihole residue.
Interaction with phonons in the insulating phase can easily reduce the residue
by 20-30\%. Stability of the disordered spiral in the insulating phase is due
to localization of holes. The $E_{cross}$ in this case is~\cite{luscher07b}
\begin{equation}
\label{ecross1}
E_{cross}\approx c\sqrt{\frac{3}{4}}\frac{Q^2}{\kappa} \ ,
\end{equation} 
where $\kappa$ is the inverse localization length.
It is worth noting that Eq. (\ref{ecross1}) has been derived in 
Ref.~\cite{luscher07b} assuming that the binding energy of a hole trapped
by Sr ion is larger than the magnon energy. The binding energy is about
10-15meV. Therefore, strictly speaking, Eq. (\ref{ecross1}) is applicable only
at $x=0.025$ since at larger $x$ the energy $E_{cross}$ is getting too big.
Nevertheless, we can try to apply (\ref{ecross1}) to the data
at  $x \leq 0.055$.
Fitting the data from Table~I  we find values of $\kappa$,
$x=0.025$: $\kappa=0.55\pm 0.2$, $x=0.04$: $\kappa=0.65\pm 0.2$, and 
$x=0.05$: $\kappa=0.75\pm 0.2$.
So, there is a hint for a weak doping dependence of the inverse localization
length $\kappa$.
Most likely the dependence is just an imitation of the binding energy
correction to formula (\ref{ecross1}). On the other hand, a weak
dependence of the localization length on doping is quite possible.
The above values agree reasonably well with the value 
$\kappa \approx 0.4$ that follows from the analysis of the variable range
hopping conductivity at a very small doping ($x = 0.002$), see 
Ref.~\cite{chen95}.

Near $x=0.12$ certain La-based materials in LTT phase develop a strongly enhanced 
static incommensurate magnetic order accompanied by a small lattice deformation at 
the second order harmonics~\cite{Tranquada95,Tranquada96,Fujita04}, see also
Ref.~\cite{tranquada05} for a review. 
The measured static magnetic moment $\sim 0.1\mu_b$ is substantially larger than the
value that follows from the present theory
(the unity in the vertical scale in Fig.~\ref{Zout}B corresponds to the magnetic
moment $0.6\mu_B$).
 There are also  experimental
indications that the spin structure in this case is close to 
collinear~\cite{christensen}.
We strongly believe that physics of these materials is somehow related to
mechanisms considered in the present paper. On the other hand, it is clear
that in this case there are some additional effects that are not accounted for by the
present theory.

The present theory qualitatively explains the directional nematic state discovered in 
underdoped YBCO at doping $x \approx 0.08$~\cite{hinkov07b}.
For a quantitative comparison one needs to analyse the two layer situation.
This analysis has to include an explanation of a smaller incommensurability
compared to that observed in the single layer LSCO.

In the present work we did not account for the superconducting pairing.
The point is that at low doping  the pairing practically does not influence
magnetic excitations.
A different question is how the spiral and the corresponding 
magnetic excitations influence the
superconducting pairing.
The spin-wave exchange mechanism for pairing of holons was suggested 
in  Refs.~\cite{flambaum,belinicher}.
The mechanism is always working as soon as a short range antiferromagnetic order
exists in the system. So the superconductivity peacefully coexists with spin 
spirals~\cite{sushkov04}.
Moreover, we understand now that the pairing in the spiral state is strongly enhanced 
by closeness to the N\'eel state instability driven by the parameter $\lambda$.
The enhancement will be considered elsewhere.

In conclusion, 
using the low-energy effective field theory we have considered
the 2D $t$-$J$ model in the limit of  small doping.
Quantitatively this consideration is relevant to underdoped single layer
cuprates.
We have derived the incommensurate spin structure (static and/or dynamic),
calculated spectra of magnetic excitations (Figs. \ref{gfi}, \ref{spe},
\ref{fig:outofplane}), and considered the quantum
phase transition to the directional nematic spin-liquid phase.
The spin wave pseudogap is opened in the spin-liquid phase, the q-integrated
spectral density in this case is shown in Fig.\ref{Inematic}.

\acknowledgments
We are very grateful to G.~Khaliullin, B.~Keimer, J.~Sirker and A.~Katanin,
for valuable discussions. We are also very grateful to B.~Keimer and K.~Yamada
for communicating their results prior to publication.
A.~I.~M. gratefully acknowledges the School of Physics at the University of New
South Wales for warm hospitality and financial support during his visit.
This work was supported in part by the Australian Research Council.


\begin{thebibliography}{99}
\bibitem{keimer92} B.~Keimer, A.~Aharony, A.~Auerbach, R.~J. Birgeneau, A.~Cassanho, Y.~Endoh, R.~W. Erwin, M.~A. Kastner, and G.~Shirane, \prb {\bf 45}, 7430 (1992).
\bibitem{kastner98} M.~A. Kastner, R.~J. Birgeneau, G.~Shirane, and. Y.~Endoh, 
\rmp {\bf 70}, 897 (1998).
\bibitem{matsuda02} M.~Matsuda, M.~Fujita, K.~Yamada, R.~J. Birgeneau, Y.~Endoh, and G.~Shirane, \prb {\bf 65}, 134515 (2002).
\bibitem{wakimoto99} S.~Wakimoto, G.~Shirane, Y.~Endoh, K.~Hirota, S.~Ueki, K.~Yamada, R.~J. Birgeneau, M.~A. Kastner, Y.~S. Lee, P.~M. Gehring, and S.~H. Lee, \prb {\bf 60}, R769 (1999).
\bibitem{matsuda00} M.~Matsuda, M.~Fujita, K.~Yamada, R.~J. Birgeneau, M.~A. Kastner, H.~Hiraka, Y.~ Endoh, S.~Wakimoto, and G.~Shirane, \prb {\bf 62}, 9148 (2000).
\bibitem{fujita02} M.~Fujita, K.~Yamada, H.~Hiraka, P.~M. Gehring, S.~H. Lee, S.~Wakimoto, and G. ~Shirane, \prb {\bf 65}, 064505 (2002).
\bibitem{yamada98} K.~Yamada, C.~H. Lee, K.~Kurahashi, J.~Wada, S.~Wakimoto, S.~Ueki, H.~Kimura, Y.~Endoh, S.~Hosoya, G.~Shirane, R.~J. Birgeneau, M.~Greven, M.~A. Kastner, and Y.~J. Kim, \prb {\bf 57}, 6165 (1998).
\bibitem{yamada07} K.~Yamada, private communication.
\bibitem{Tranquada95} J.~M.~Tranquada {\it et al}, Nature {\bf 375}, 561 (1995). 
\bibitem{Tranquada96} J.~M.~Tranquada {\it et al}, Phys. Rev. B {\bf 54}, 7489 (1996).
\bibitem{Fujita04} M.~Fujita {\it et al}, Phys. Rev. B {\bf 66}, 184503 (2002);
{\bf 70}, 104517 (2004).
\bibitem{tranquada05} J.~M. Tranquada, cond-mat/0512115.
\bibitem{bourges00} P.~Bourges, Y.~Sidis, H.~F. Fong, L.~P. Regnault, J.~Bossy, A.~Ivanov, and B.~ Keimer, Science {\bf 288}, 1234 (2000).
\bibitem{fong00} H.~F. Fong, P.~Bourges, Y.~Sidis, L.~P. Regnault, J.~Bossy, A.~Ivanov, D.~L. Milius, I.~A. Aksay, and B.~Keimer, \prb {\bf 61}, 14773 (2000).
\bibitem{mook02} H.~A. Mook, Pengcheng Dai, F.~Dogan, \prl {\bf 88}, 097004 (2002).
\bibitem{hayden04} S.~M. Hayden, H.~A. Mook, Pengcheng Dai, T.~G. Perring, and F.~Do\v{g}an, Nature {\bf 429}, 53 (2004)
\bibitem{hinkov04} V.~Hinkov, S.~Pailh\`es, P.~Bourges, Y.~Sidis, A.~Ivanov, A.~Kulakov, C.~T. Lin, D.~P. Chen, C.~Bernhard, and B.~Keimer, Nature {\bf 430}, 650 (2004)
\bibitem{stock06} C.~Stock, W.~J.~L. Buyers, Z.~Yamani, C.~L. Broholm, J.-H.~Chung, Z.~Tun, R.~Liang, D.~Bonn, W.~N. Hardy, and R.~J. Birgeneau, \prb {\bf 73}, 100504 (2006).
\bibitem{hinkov07a}
V.~Hinkov, P.~Bourges, S.~Pailhes, Y.~Sidis, A.~Ivanov, C.~D.~Frost, 
T.~G.~Perring, C.~T.~Lin, D.~P.~Chen, B.~Keimer,
Nature Physics {\bf 3}, 780 (2007).
\bibitem{hinkov07b} V.~Hinkov, D.~Haug, B.~Fauque, P.~Bourges, Y.~Sidis, 
A.~Ivanov, C.~Bernhard, C.~T.~Lin, B.~Keimer, ??? (2007).
\bibitem{PWA} P. W. Anderson, Science \textbf{235}, 1196 (1987).
\bibitem{Em} V. J. Emery, Phys. Rev. Lett. \textbf{58}, 2794 (1987).
\bibitem{ZR} F. C. Zhang and T. M. Rice, Phys. Rev. B \textbf{37}, R3759
(1988).
\bibitem{tokura90} Y.~Tokura, S.~Koshihara, T.~Arima, H.~Takagi, S.~Ishibashi, T.~Ido, and S.~Uchida, \prb {\bf 41}, R11657 (1990).
\bibitem{andersen95} O.~K. Andersen, A.~I. Liechtenstein, O.~Jepsen, and
F.~Paulsen, J. Phys. Chem. Solids \textbf{56}, 1573 (1995); E.~Pavarini,
I.~Dasgupta, T.~Saha-Dasgupta, O.~Jepsen, and O.~K. Andersen ,\prl {\bf 87}
047003 (2001).
\bibitem{shraiman88} B.~I. Shraiman and E.~D. Siggia, 
\prl {\bf 61}, 467 (1988);
\prl {\bf 62}, 1564 (1989); \prb {\bf 42}, 2485 (1990).
\bibitem{Dom} T. Dombre, J. Phys. (France) {\bf 51}, 847 (1990).
\bibitem{IF} J. Igarashi and P. Fulde, Phys. Rev. B {\bf 45},
10419 (1992).
\bibitem{CM} A. V. Chubukov and K. A. Musaelian, Rev. B {\bf 51}, 12605 (1995).
\bibitem{hasselmann04} N.~Hasselmann, A.~H. Castro Neto, and C.~Morais Smith, \prb {\bf 69}, 014424 (2004).
\bibitem{sushkov04} O.~P. Sushkov and V.~N. Kotov, \prb {\bf 70}, 024503 (2004).
\bibitem{juricic04} V.~Juricic, L.~Benfatto, A.~O. Caldeira, and C.~Morais Smith, \prl {\bf 92}, 137202 (2004).
\bibitem{sushkov05} O.~P. Sushkov and V.~N. Kotov, \prl {\bf 94}, 097005 (2005).
\bibitem{lindgard05} P.-A.~Lindg{\aa}rd, \prl {\bf 95}, 217001 (2005).
\bibitem{juricic06} V.~Juricic, M.~B. Silva-Neto, and C.~Morais Smith, \prl {\bf 96}, 077004 (2006).
\bibitem{luscher06} A.~L\"uscher, G.~Misguich, A.~I. Milstein, and O.~P. Sushkov, \prb {\bf 73}, 085122 (2006).
\bibitem{luscher07} A.~L\"uscher, A.~I. Milstein, and O.~P. Sushkov, \prl {\bf 98}, 037001 (2007).
\bibitem{luscher07b} A.~L\"uscher, A.~I. Milstein, and O.~P. Sushkov, \prb {\bf 75}, 235120 (2007).
\bibitem{SdH}  N.~Doiron-Leyraud, C.~Proust, D.~LeBoeuf, J.~Levallois, 
J.-B.~Bonnemaison, R.~Liang, D.~A.~Bonn, W.~N.~Hardy,  and  L.~Taillefer,
Nature {\bf 447}, 565 (2007).
\bibitem{dagotto94} E.~Dagotto, \rmp {\bf 66}, 763 (1994).
\bibitem{wiegman88} P.~Wiegman, \prl, {\bf 60}, 821 (1988).
\bibitem{wen89} X.~G. Wen, \prb {\bf 39}, 7223 (1989).
\bibitem{W}
F.~Kampfer, M.~Moser, and U.-J.~Wiese, Nucl. Phys. B {\bf 729}, 317 (2005);
C.~Brugger, F.~Kampfer, M.~Moser, M.~Pepe, and U.-J.~Wiese,
Phys. Rev. B {\bf 74}, 224432 (2006). 
\bibitem{SZ} R.~R.~P.~Singh, Phys. Rev. B {\bf 39}, 9760 (1989);
Zheng Weihong, J.~Oitmaa, and C.~J.~Hamer, Phys. Rev B
{\bf 43}, 8321 (1991).
\bibitem{takahashi} M.~Takahashi, 
         Phys.\ Rev.\ Lett. {\bf 58}, 168 (1987);
         Phys.\ Rev.\ B {\bf 40}, 2494 (1989).
\bibitem{chandra} P.~Chandra, P.~Coleman, and A.~I.~Larkin, J. Phys.: Condens.
Matter {\bf 2}, 7933 (1990).
\bibitem{chen95}
C.~Y.~Chen, E.~C.~Branlund, ChinSung Bae, K.~Yang,
M.~A.~Kastner, A.~Cassanho, and R.~J.~Birgeneau, Phys. Rev. B 51, 3671 (1995).
\bibitem{christensen}
N.~B.~Christensen, H.~M.~Ronnow, J.~Mesot, R.~A.~Ewings, N.~Momono, M.~Oda, 
M.~Ido, M.~Enderle, D.~F.~McMorrow, A.~T.~Boothroyd,
Phys. Rev. Lett. {\bf 98}, 197003 (2007).
\bibitem{flambaum}
V.~V.~Flambaum, M.~Yu.~Kuchiev, and O.~P.~Sushkov,
Physica C {\bf 227}, 267-278 (1994).
\bibitem{belinicher}
V.~Belinicher, A.~Chernyshev, A.~Dotsenko, and O.~P.~Sushkov,
Phys. Rev. B, {\bf 51}, 6076-6084. (1995).

 
\end{thebibliography}
\end{document}